\documentclass[preprint,preprintnumbers]{revtex4}
\begin{document}
\title {Tunable Self-assembly of Au-nanoparticles with Near-Perfect Monodispersity}
\author { S. Chattopadhyay, A. Datta}
\affiliation {Surface Physics Division, Saha Institute of Nuclear Physics,
1/AF, Bidhannagar, Calcutta 700 064. India.}
\begin{abstract}
Au nanoparticles with near perfect monodispersity in shape and size
self-assemble, very slowly and at room temperature, on polystyrene
films for film thickness $\leq 4R_{g}$, $R_{g}$ being unperturbed
polymer gyration radius. Nanoparticle shape and size can be tuned,
without sacrificing monodispersity, by changing polymer film
thickness. Self-assembly is caused by enhanced two-dimensional
diffusion of sputter-deposited Au clusters on a distribution of
surface energy variations, sharply defined and with tunable
dimensions, corresponding to low cohesion between gyration spheres
on film surface. Lowering of cohesion and enhanced surface diffusion
are clearly related to confinement-induced gyration sphere layering
along film depth.
\end{abstract}
\maketitle

The fundamental aspect of nanoparticles is the emergence of totally
new structures and properties that are crucially dependent on the
size and shape of the nanoparticles \cite{rmp05, rpp01}. Hence a
system of nanoparticles has well-defined structure and properties
only when it is monodisperse, i.e., when all nanoparticles in the
system have the same size and shape. For metal nanoparticles this
means that each particle has a sharply defined number of metal atoms
arranged in a specific way, which, when changed to some other
arrangement, gives a new well-defined system of nanoparticles.
Nanoparticles are thus different from both the bulk solid, where the
number of atoms is indefinite, and the molecule, where it is
unchangeable. The prevalent methods of producing nanoparticles are
thus either $`$top-down' or $`$bottom-up'. In the first, nanosized
pieces are broken off the bulk \cite{sci03, nat01, jap00} and
deposited on a substrate. In the second, nanoclusters of specific
atoms are assembled using specific chemical interactions
\cite{sci02, apl01, jcs94}. However, both methods have inherent
problems. In the first, or physical, method it is easy to tune the
number of atoms and their arrangements in the nanoparticle by
changing the energy of collision and / or the crystal face of the
bulk  but the system produced is generally not monodisperse
\cite{prb05}. This is due to the inherent energy-width of the
projectile-solid interactions that determine the morphology of
nanoparticles grown from these projectile clusters\cite{nimb97}. In
the second, or chemical, method remarkable monodispersity is
achieved due to specificity of chemical reaction \cite{prb02,prl00}
but this very specificity makes changing the nanoparticle shape and
size nearly impossible and generalized chemical methods for growing
nanoparticles is the focus of recent research\cite{nat05, nat00}.

A nanoparticle can be viewed also as a cluster of atoms of a
specific number and arrangement, which is in metastable equilibrium
\cite{rmp05}. In other words, nanoparticles correspond to local
minima in free energy as functions of the cluster shape and size
and, if suitable $`$potential wells' are provided, atomic clusters
will most probably occupy them, thus forming well-defined
nanoparticles. Changing the depth and width of such $`$potential
wells' would then change the shape and size of the nanoparticle and
would, at each case, sharply choose a particular shape and size,
preserving monodispersity. This concept forms the basis of another
mode of producing nanoparticles, that of \emph{tunable
self-assembly}.

In this communication we have presented a realization of such
tunable self-assembly of Au nanoparticles sputter-deposited on a
spin-coated polystyrene (\emph{PS}) film on amorphous quartz. We
have found that, reducing thickness of the \emph{PS} film reduces
adhesion of the initially deposited Au, resulting in an enhanced
ambient surface diffusion of initially formed Au clusters - a slow
two-dimensional Brownian motion over a surface layer having
effective viscosity at room temperature orders of magnitude lower
than bulk PS viscosity near glass transition temperature ($T_{G}
\simeq$ 100$^{\circ}$C)\cite{macro99}. At the same time we find that
the film makes the clusters form larger nanoparticles with
near-perfect monodispersity in size and shape that can be tuned by
changing film thickness within a range, without sacrificing the
monodispersity. We have found that the formation of nanoparticles
with fixed size and shape is achieved by trapping the clusters in
sharply defined surface energy gradients ($`$wells') with fixed
dimensions, appearing on \emph{PS} film surface whose dimensions
decrease with increase in film thickness.

Polystyrene (mol.~wt.~$M\simeq 560900$, unperturbed radius of
gyration $R_{g}=0.272M^{\frac{1}{2}}\simeq$ 20.4nm) \cite{book1} was
spin-coated on fused quartz from toluene solutions to form films
with air/film and film/substrate interfacial roughness $\sim$ 0.6nm.
The thickness was controlled by a combination of rotation speed and
solution concentration and varied from 40 nm ($\simeq 2R_{g}$) to
180 nm ($\simeq 9R_{g}$). A series of these \emph{PS} films was used
as the pristine (\emph{pPS}) samples while Au was D.C.
sputter-coated for 10s \cite{note1} on a second identical series
forming the \emph{AuPS} samples.

Figure 1((a)-(c)) show the topographic images of Au-nanoclusters,
obtained from tapping mode Atomic Force Microscopy (AFM),
immediately after sputter deposition on a \emph{PS} film of
thickness $d \simeq 2R_{g}$ (Figure1(a)), and after two months
(Figure 1(b)) and six months post-deposition (Figure 1(c)) all
taken at room temperature. While the first image shows no
cluster-like features, combination of results from the plasmon
peak in transmission optical spectra \cite{book2} and the electron
density profile (EDP) (top inset, Figure 1(a)) extracted from
x-ray reflectivity (bottom inset, Figure 1(a))\cite{pac02}
confirms the presence of a layer of spherical Au clusters with
diameter $\simeq$ 3 nm only on top of the \emph{PS} films (top
inset and bottom inset of Figure1(a)are not shown). The image
taken after two months, on the other hand, shows presence of
larger and ellipsoidal nanoparticles as indicated by in-plane
semi-axes, \emph{a} and \emph{b} and out-of-plane semi-axis
\emph{c}(\emph{a} $\approx$ \emph{b} $>$ \emph{c}) (Table 1),
whereas in Figure 1(c) there is almost no change in these
dimensions suggesting that growth of the particles has stopped.
From the values of \emph{a}, \emph{b} and \emph{c} over time
(Table 1) it is clear that initial small and spherical
nanoparticles are coalescing, predominantly in-plane and at room
temperature, to form larger and ellipsoidal nanoparticles.

Figures 1(d) and 1(e) show the images taken just after and two
months after Au deposition, respectively, on $d \simeq 4R_{g}$
\emph{PS} film and Figures 1(f) and 1(g) show the same time sequence
for the $d\simeq 9R_{g}$ film. Au was deposited and kept under
conditions identical with $d\simeq 2R_{g}$ film. Coalescence of
particles is again observed in the $4R_{g}$ thick film with some
modifications (Table 1) but there is almost no coalescence in the
$9R_{g}$ thick \emph{PS} film.

The lower panel of Figure 1 show the reflectivity profiles of
\emph{pPs} films of $d \simeq 2R_{g}$ (Figure 1(h)), $d \simeq
4R_{g}$ (Figure 1(i)) and $d \simeq 6R_{g}$ or 120nm (Figure1(j)).
In each case the reflectivity data is shown by open circles while
the calculated best fit is shown by solid line and the extracted
EDP is shown in inset. As expected \cite{prb}, below $d\simeq
4R_{g}$, \emph{PS} molecules form layers along film thickness with
periodicity $\sim R_{g}$. The key point is to note that this layer
formation is strongly correlated with surface coalescence of Au
nanoparticles. (For Figure 1(h), (i), (j) see \cite{prb}: Fig 3(a)
and 3(b))

Figure 2 presents the monodispersity of the self-assembled
nanoparticles. Figure 2(a) and 2(b) show, respectively, topographic
and phase images of Au-nanoparticles formed on the 2$R_{g}$ thick
\emph{PS} film two months after deposition, while Figure 2(c)
presents (in filled circles) $N$, the number of nanoparticles with a
particular value of $a (\simeq b)$ as a function of $a$. We have
averaged over four 2$\mu$m $\times$ 2$\mu$m scans. The size
distribution in our case is completely different and much sharper
than the log-normal distribution of sizes of nanoparticles formed by
random coalescence \cite{jap92}. This precludes any random
coalescence mechanism such as the Vollmer-Weber growth \cite{prl05}
and it strongly suggests presence of potential wells with fixed
depth and width that act as size-selective traps for the
Au-nanoparticles to fall into and form the larger nanoparticles. We
have also found that this monodispersity is not disturbed for $d\leq
4R_{g}$.

Figures 3(a), 3(b), and 3(c) show the topographical and Figures
3(d), 3(e) and 3(f) show the phase image of \emph{pPS} films of
$d=2R_{g}$, $4R_{g}$ and $9R_{g}$, respectively. All the
topographical images look the same, having roughly spherical
features with an average diameter of $R_{g}$, as seen previously
\cite {prb} and shown, for example, in the line profile (inset,
Figure 3(b)(not shown)), clearly indicating the presence of
gyration spheres in \emph{pPS} films at all thickness with a size
modification probably due to entanglement and substrate effects.
On the other hand the phase images exhibit larger changes in
phase-shifts between adjacent $`$spheres' as the thickness is
reduced, implying a larger change in energy being dissipated by
the AFM tip in going over from one sphere to another\cite{apl97}.
We have estimated this average energy dissipated by the tip over
the film surfaces, using the expression \cite{apl98}
\begin{equation}
\sin\phi=(\frac{\omega}{\omega_{0}}\frac{A}{A_{0}})+\frac{QE_{D}}{\pi
k A A_{0}}
\end{equation}
where $\phi$ is the phase-shift, $\omega$ ($\omega_{0}$) is the
working (resonance) frequency, $A$ ($ A_{0}$) is the setpoint (free)
amplitude, Q is the quality factor,and $k$ is the cantilever spring
constant while $E_{D}$ is the energy dissipation.

The average energy dissipated per unit tip-sample contact area is
then given by \cite{book1}
\begin{equation}
W_{D}= \frac{E_{D}}{4 \pi a_{Si}r_{c}}
\end{equation}
where $r_{c}$ = radius of tip-curvature, $a_{Si}$ = diameter of Si
atom of tip. The surface energy ($\gamma_{PS}$) of the polymer film
is then given in terms of the interfacial energy ($\gamma_{Si-PS}$)
and the Si surface energy ($\gamma_{Si}$) as \cite{book1}
\begin{equation}
\gamma_{Si-PS} = -\frac{W_{D}}{2} = (\gamma_{Si}^{\frac{1}{2}} -
\gamma_{PS}^{\frac{1}{2}})^2
\end{equation}
We have obtained average $\gamma_{PS}\simeq 30mJm^{-2}$,
consistent with bulk polystyrene values ($\simeq
33mJm^{-2}$)\cite{book1}. In Figures 3(g), 3(h), 3(i), we have
shown $\Delta\gamma_{PS}(x,y)$, the surface energy relative to top
of the gyration spheres, derived from Figure 3(d), 3(e) and 3(f),
respectively (Figure 3(a), (b), (c), (d), (e), (f), (g), (h), (i)
are not shown, Figure 3(j) corresponds to them). Variation in
surface energy, i.e. magnitude of $\Delta\gamma_{PS}(x,y)$ between
adjacent gyration spheres increases as film thickness is reduced,
indicating the $`$disentanglement' of the spheres due to decrease
in cohesion. Typical line profiles ($\Delta\gamma_{PS}(x)$) across
such variations presented in Figure 3(j) show a decrease in
surface energy on top of spheres relative to their contact region.
For non-H-bonding solids and liquids the Hamaker constant, $A_H
\approx 2.1\times10^{-21}\gamma$ ($\gamma$ in mJm$^{-2}$ and $A_H$
is in Joules). The decrease in Hamaker constant ($\Delta A_H$)
between the top and bottom of the line profiles, i.e., between
contact region of adjacent gyration spheres and top of spheres,
obtained from the above expression, are given in Table 1. They
match values of the decrease in Hamaker constant between molecular
layers, formed by confinement, along depth, in films of the
corresponding thickness \cite{prb}, also presented in Table 1.
Thus confinement lowers cohesion between adjacent gyration
spheres, ie, molecules of \emph{PS}, both parallel and
perpendicular to the direction of confinement. The decreasing
variation in surface energy with increasing film thickness is
clearly consistent with the decreasing dimensions of the coalesced
nanoparticles (Table 1) and final stopping of coalescence. This
suggests that the driving mechanism of the coalescence is the
sharply defined in-plane gradients in surface energy.

In the case of lowering of cohesion across the film thickness, we
showed that this lowering is associated with breakdown of
intermolecular excitonic coupling through J-interaction between
adjacent benzene rings \cite{prb}. If we assume that the same
mechanism to be active along the film plane, we would expect
consequent changes in the in-plane properties such as
visco-elasticity.

$T_{G}$ of bulk \emph{PS}, or even its surface ($\simeq$
77$^{\circ}$C)\cite{macro99}, is much above room temperature. The
fact that Au nanoparticles are still diffusing on \emph{PS} surface,
albeit slowly, attests to low Au-\emph{PS} adhesion \cite{pre02}.
Therefore two-dimensional Brownian motion is the preferred mechanism
of diffusion \cite{jpcb05}. It is also known that, near $T_{G}$,
there is a $`$liquid-like' surface layer of thickness $\sim 4nm$ for
polymers in general \cite{epj04} on which surface diffusion of
clusters has been found to be orders of magnitude faster than bulk
diffusion \cite{prl05}. If we assume a two-dimensional Brownian
motion even at ambient conditions, for the initially deposited Au
clusters, we can estimate the effective viscosity of the $`$liquid'
surface layer of the film at room temperature. We have used the
Smoluchowski relation \cite{book3}
\begin{equation}
\eta=\frac{4 k_{B}Tt}{6 \pi \langle r ^{2}\rangle a_{D}}
\end{equation}
where $\langle r ^{2}\rangle^{\frac{1}{2}}$ =   distance traversed
by diffusing Au cluster in time $t$, $a_{D}$ = cluster radius,
$\eta$ = effective viscosity of surface $`$liquid' layer at
temperature $T$  and $k_{B}$ = Boltzmann constant. If we take
$2a_{D} \simeq$ 3 nm, $\langle r^{2}\rangle^{\frac{1}{2}}$ =
radius of \emph{pPS} gyration sphere measured from Figure 3(FWHM
of line profile) $\sim R_{g}/2$ = 9 nm, we arrive at $\eta$ = 3.71
$\times$ $10^{11}$ Poise for $d \simeq 2R_{g}$ at room
temperature, which is two order of magnitude less than
$\eta_{bulk}$ for \emph{PS} (at $T_{G}$) \cite {book4} (Table 1).
For $d\simeq 4R_{g}$ the value is of the same order but for
$d\simeq 9R_{g}$, as there is almost no coalescence, the surface
behaves essentially like bulk \emph{PS} at room temperature. The
large decrease in surface viscosity due to confinement strongly
suggests a compositional change on the surface of gyration spheres
of \emph{PS}, accounting for both the lowering of cohesion and
Au-\emph{PS} adhesion (causing enhanced effective viscosity). In
Figure 3(k) we have shown plasmon peaks of Au nanoparticles, just
after deposition, on \emph{PS} with $d \simeq 2R_{g}$ and $4R_{g}$
(Figure 3(k) is not shown). Under identical deposition conditions,
the peak at 580 nm, corresponding to individual nanoparticle, is
seen to be reduced while the other (non-gaussian) peak,
corresponding to inter-nanoparticle coupling \cite{lang04}, is
absent for the thinner film. This indicates that lesser amount of
Au adhere to the surface of the thinner film, to begin with
\cite{prl99} and it also explains why fewer nanoparticles are
formed in this case.

No chemical reactions could possibly have taken place while
thinning. Also we know that thinning breaks the intermolecular
benzene-benzene connection \cite{prb}. We then suggest that a
probable mechanism for all the observed effects is a rearrangement
on surfaces of the gyration spheres \cite{jpc02} whereby the
proportion of benzene rings on these surfaces is reduced with
respect to that in bulk \emph{PS}. The lower Au-methylene adhesion,
as compared to Au-benzene adhesion causes faster diffusion of the Au
clusters. Due to lack of benzene rings, the J-coupling between
spheres also reduces, lowering cohesion between adjacent spheres and
forming gradients in surface energy that trap the diffusing
nanoclusters to form monodisperse nanoparticles.

The confinement-induced tuning of surface energy described above
is more generalized than chemically controlling surface energy
\cite {sci05}. The biggest advantage of using polymers in the
growth of nanoparticles, which is essentially a phenomenon at
mesoscopic lengthscales, is that the molecules of a polymer are
themselves objects at such lengthscales. Thus molecular
rearrangements at polymer surfaces can, as in our case, be
employed to direct nanoparticle growth. However, such
rearrangements can also cause very interesting changes in other
surface properties at mesoscales, eg, optical polarization as a
function of film thickness. Work in these directions is underway.
(Some figures are described but not shown here due to insufficient
space in arXiv).

\begin{figure}
{\bf Figure captions} \caption{(a)-(g): Topographic images of
($500nm\times 500nm$) Atomic Force Microscopic (AFM) scans in
tapping mode (intermittent contact) of Au sputter deposited on
polystyrene (\emph{PS}) films. Thickness, in units of $R_{g}$=
unperturbed \emph{PS} gyration radius ($\simeq$ 20.4nm), and time
lapsed after Au deposition given in pairs. (a) 2$R_{g}$, immediate;
(b) 2$R_{g}$, 2 months; (c) 2$R_{g}$, 6 months; (d) 4$R_{g}$,
immediate; (e) 4$R_{g}$, 2 months; (f) 9$R_{g}$, immediate; (g)
9$R_{g}$, 2 months. Observed (open red circles) and best fit (black
line) x-ray reflectivity profile and Electron Density Profile (EDP)
extracted from fit for the corresponding Au-\emph{PS} film are
shown, bottom and top insets of (a), respectively. (h)-(j):
Reflectivity profiles, observed (open red circles) and calculated
(black lines) with EDP's extracted from calculated profiles (insets)
for pristine \emph{PS} films with thickness (i) 2$R_{g}$, (j)
4$R_{g}$, (k) 9$R_{g}$.}

\caption{(a) Topographic and (b) Phase images of ($2\mu m \times
2\mu m$) AFM scans of Au nanoparticles formed on 2$R_{g}$ thick PS
film 2 months after Au deposition. (c) \emph{N} vs \emph{a} plot (in
filled circles), $a (\simeq b)$ is the in-plane semi-major axis of
the nanoparticles in (a) and (b) and $N$ is the number of
nanoparticles with particular value of $a$.}

\caption{(a)-(c): Topographic and (d)-(f): Phase images of ($500nm
\times 500nm$) AFM scans of pristine \emph{PS} films with thickness
$2R_{g}$ ((a) and (d)), $4R_{g}$ ((b) and (e)) and $9R_{g}$ ((c) and
(f)). Inset (b): line profile along a gyration sphere. (g)-(i):
In-plane maps of $\Delta\gamma_{PS}$, surface energy relative to top
of the gyration spheres (shown in (a)-(c), respectively, refer text
for details). (j): $\Delta\gamma_{PS}$, plotted along typical lines
in (a), (b) and (c), across adjacent gyration spheres given by
curves 1, 2 and 3, respectively. (k): Plasmon peaks in optical
transmission spectra of Au-\emph{PS} films, just after Au deposition
for \emph{PS} thickness (1) 2$R_{g}$ and (2) 4$R_{g}$.}

\end{figure}

\newpage
\begin{table}
  \centering
  \caption{Structure of Au nanoparticles and visco-elasticity of polystyrene films}\label{rt}
  \begin{tabular}{|c|c|c|c|c|c|c|c|c|c|c|c|}
  \hline\hline
  \multicolumn{1}{|c|}{\emph{PS}} & \multicolumn{6}{|c|}{Au nanoparticle} &
 In-plane & Interlayer & Effective  &
  Bulk  \\
  film & \multicolumn{6}{|c|}{dimension (nm)}& Hamaker constant  & Hamaker constant
  & surface& viscosity \\ \cline{2-7}
  \multicolumn{1}{|c|}{thickness} &  \multicolumn{3}{|c|}{as deposited\footnotemark[1]} &
    \multicolumn{3}{c|}{after 2 months\footnotemark[2]} &variation ($\triangle A_H$)
    & variation ($\triangle A_H$) \footnotemark[3]& viscosity of \emph{PS} & of \emph{PS}\footnotemark[4] \\ \cline{2-7}
  (nm)& \hspace {.1in}\textbf{\emph{a}} \hspace {.1in}& \hspace {.1in}\textbf{\emph{b}} \hspace {.1in}& \textbf{\emph{c}}
  & \hspace {.1in}\textbf{\emph{a}}\hspace {.1in} &
   \hspace {.1in}\textbf{\emph{b}}\hspace {.1in} & \textbf{\emph{c}} & (meV) & (meV) & $\eta$, (Poise) & $\eta_{bulk}, $(Poise) \\ \hline
   $2R_{g}$& 3 & 3 & 3 & 26 & 26 & 6 & 63 & 97& $\sim$ 3.71$\times 10^{11}$ & \\ \cline{1-10}
   $4R_{g}$ & 3 & 3 & 3 & 10 & 10 & 3 &  15 & 28& $\sim$ 3.71$\times 10^{11}$ &$\sim 10^{13}$ \\
   \cline{1-10}
   $9R_{g}$& 3 & 3 & 3 & 3 & 3 & 3 & 8 & $\sim$ 0& -&  \\ \hline
\end{tabular}
\footnotetext[1]{From plasmon spectra and x-ray reflectivity}
\footnotetext[2]{From spectra, AFM and x-ray reflectivity}
\footnotetext[3]{From \cite{prb}} \footnotetext[4]{Above $T_G$, from
\cite{book4}}
\end{table}


\begin{thebibliography}{28}
\bibitem{note1} Au was sputtered on PS films of different thicknesses
by a  DC magnetron coater (Pfeiffer) for  10s at 25W, 10 sccm Ar gas
flow, $4.7\times10^{-3}$ mbar pressure (base pressure =
$8.6\times10^{-6}$ mbar).
\end{thebibliography}
\end{document}